# Enhanced quantum efficiency and reduction of reflection for MSM photodetectors with nano-structured surface

Ekaterina Ponizovskaya Devine
*Electrical and Computer Engineering*
*UC Davis*
Davis, California, USA
eponizovskayadevine@ucdavis.edu

Hilal Cansizoglu
*Electrical and Computer Engineering*
*UC Davis*
Davis, California, USA
hcansizoglu@ucdavis.edu

Yang Gao
*Electrical and Computer Engineering*
*UC Davis*
Davis,Califirnia, USA
yangao@ucdavis.edu

Cesar Perez
*Electrical and Computer Engineering*
*UC Davis*
Davis,Califirnia, USA
cperez@ucdavis.edu

Toshishige Yamada
*W&WSens Device Inc*
*Los Altos, California, USA*
t.yamada@ieee.org

Aly F. Elrefaie
*W&WSens Device Inc*
*Los Altos, California, USA*
aelrefaie@ucdavis.edu

M.Saif Islam
*Electrical and Computer Engineering*
*UC Davis*
Davis,California, USA
sislam@ucdavis.edu

Shih-Yuan Wang
*W&WSens Device Inc*
*Los Altos, California, USA*
wang006@gmail.com

*Abstract*— The photon trapping nano-structures help to enhance quantum efficiency and reduce reflection for MSM photodetector that allows fast Si photodetectors at wavelength 800-950nm. The nanostructure consist of micro holes reduces reflection and bends normally incident light into the lateral modes in the absorbing layer.

## I. Introduction

The Si MSM photodetectors (PDs) at the short-reach multimode datacom wavelengths of 840-860 nm and at a short wavelength division multiplexing (SWDM) band of 850-950 nm could facilitate the integration [1] and reduce the cost. Our study [2,3] showed theoretically and experimentally that a nano- or micro-structure can increase the Si PDs quantum efficiency (QE) for pin diodes so that less than 2 microns intrinsic (i) layer can show the QE>50% with data rate 20 Gb/s for wavelength 800-950nm. In this study we use nanoholes to reduce the reflection and increase QE in MSM PDs. In order to increase the speed, the shorter distance between electrodes can be used. In a flat device it can cause a significant reflection. However, we have shown that the holes in the Si surface between electrodes can drastically reduce reflection.

## II. Design and Simulations

The designed PDs have 2D periodic nano-holes in the shape of inverted pyramids Fig. 1a The view from the top (*x-y* plane) and cross-sections in the direction of light propagation (*x-z* plane) are shown in Fig. 1b,c respectively. The structure has layer of Si with thickness t on a SOI substrate and Al electrodes 300nm wide. We compared optical properties for electrodes with depth in the Si (d) equal to Si layer thickness and a half of it. The electrodes fabricated to have the same depth as Si layer are optimal for speed but in the same time they may increase losses in metal.

The Finite Difference Time Domain (FDTD) simulation shows that the flat device with the both 500nm and 1000nm electrodes designed this way produces significant, up to 80%, reflection (Fig.1d) that makes the device useless. To reduce reflection we introduce array of inverted pyramids between the electrodes with period (p) and the side of the hole (a) as it shown in Fig.1b,c. The distance between the electrodes was 1700nm. The Si thickness was taken as 1000nm and 2000nm. Two basic geometries were studied: Si thickness t=1000nm with the inverted pyramids arranged in square lattice between electrodes with period p=500nm and side a=500nm; and the thickness t=2000nm, p=1000nm, a=700nm. As it was shown in [2-4] the holes array supports a set of modes with wave vectors in *z* direction, $k_z$ and $k_c$ in *x-y* plane. Each hole couples the vertically incident light into the x-y plane. The lateral modes with larger $k_c$ can have full reflection from the bottom of the PD and they can form modes guided in the x-y plane. Fig.1d shows FDTD simulation for the E component of the optical field in x-y plane for 850nm wavelength. The lateral modes that propagates from the holes along the PDs surface and are responsible for the light trapping.

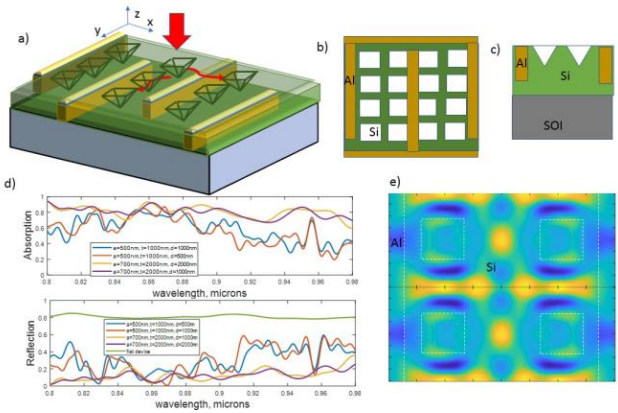

Fig.1 (a) view of MSM detector with nano-holes, (b) x-y plane; (c) x-z plane d) FDTD simulation for absorption and reflection for inverted pyramids with period (p) 500nm and 1000nm, side (a) 500nm and 700nm, in Si with thickness (t) 1micron and 2 microns, and electrodes depth (d) 500nm and 1000nm, the reflection for the electrons with d=500nm for flat device is shown at the bottom (green line) e) E field distribution around inverted pyramids in Si in x-y plane.

The leaky modes that are not fully trapped in Si can also be useful for the absorption if $Im(k_l)<\alpha$, where $k_l$ is the wave vector of the leaky mode and $\alpha$ is the Si bulk absorption. At this condition the mode can be successfully absorbed before it will leak outside the structure. The longer the light propagate into *i*-region and the better it confined in it the better is the absorption for a wider range of the wavelength. The pyramid shapes had smaller reflection due to the effect is similar to the Lambertian reflector that helps to trap light in Si [4]. The absorption and reflection are shown in Fig.2d. The simulations were averaged between both TE and TM polarizations. Lorentz model was used for Si bulk absorption and Drude model was used for metal. All the models parameters were fitted to experimental data at room temperature.

Holes shape other than inverted pyramids, for example, similar to those that was used in [5] for QE enhancement in pin photodiodes also show significant reflection reduction.

### III. Conclusion

By introducing the nano holes, the reflection was reduced from nearly 80% to less than 20%. The absorption for both electrons depth is about 80%. Some light is lost in the metal but the reduction of the reflection will drastically increase the QE. The bigger holes with a=700nm showed better anti-reflection properties. The fact that comparison of the electrodes with different thickness give similar results could be explained that most of the photons were absorbed in Si and not in the metal that allows us to expect a high QE.